\documentclass[prl,aps,twocolumn,floatfix]{revtex4-2}
\usepackage{amsmath,amssymb,amsfonts,color,graphicx,tabularx}
\usepackage[unicode=true,colorlinks=true]{hyperref}
\hypersetup{linkcolor=blue,citecolor=blue,urlcolor=blue}

\begin{document}

\title{Quantum melting of generalized electron crystal in twisted bilayer MoSe$_2$}

{\author{Qi Jun Zong$^{1}$$^{\dagger}$}
\author{Haolin Wang$^{2,3\dagger\ast}$}
\author{Qi Zhang$^{1}$$^{\dagger}$}
\author{Xinle Cheng$^{4}$$^{\dagger}$}
\author{Yangchen He$^{5}$}
\author{Qiaoling Xu$^{6,7}$}
\author{Ammon Fischer$^{4}$}
\author{Kenji Watanabe$^{8}$}
\author{Takashi Taniguchi$^{9}$}
\author{Daniel A. Rhodes$^{5}$}
\author{Lede Xian$^{6}$}
\author{Dante M. Kennes$^{4,10}$}
\author{Angel Rubio$^{10,11\ast}$}
\author{Geliang Yu$^{1,12\ast}$}
\author{Lei Wang$^{1,12\ast}$}

\affiliation{$^{1}$National Laboratory of Solid-State Microstructures, School of Physics and Collaborative Innovation Center of Advanced Microstructures, Nanjing University, Nanjing 210093, China}
\affiliation{$^{2}$School of Advanced Materials and Nanotechnology, Xidian University, Xi’an 710071, China}
\affiliation{$^{3}$Guangzhou Institute of Technology, Xidian University, Guangzhou 510080, China}
\affiliation{$^{4}$Institut für Theorie der Statistischen Physik, RWTH Aachen University and JARA-Fundamentals of Future Information Technology, Aachen 52056, Germany} 
\affiliation{$^{5}$Department of Materials Science and Engineering, University of Wisconsin, Madison, WI 53706, USA}
\affiliation{$^{6}$Songshan Lake Materials Laboratory, Dongguan, Guangdong 523808, China} 
\affiliation{$^{7}$College of Physics and Electronic Engineering, Center for Computational Sciences, Sichuan Normal University, Chengdu 610068, China} 
\affiliation{$^{8}$Research Center for Electronic and Optical Materials, National Institute for Materials Science, 1-1 Namiki, Tsukuba 305-0044, Japan} 
\affiliation{$^{9}$Research Center for Materials Nanoarchitectonics, National Institute for Materials Science, 1-1 Namiki, Tsukuba 305-0044, Japan} 
\affiliation{$^{10}$Max Planck Institute for the Structure and Dynamics of Matter,Center for Free-Electron Laser Science (CFEL), Luruper Chaussee 149, Hamburg 22761, Germany}
\affiliation{$^{11}$Center for Computational Quantum Physics, Simons Foundation Flatiron Institute, New York, NY 10010, USA}
\affiliation{$^{12}$Jiangsu Physical Science Research Center, Nanjing 210093, China}

\affiliation{$^{\dagger}$These authors contributed equally to this work.}
\affiliation{$^{\ast}$Corresponding authors, Email: hlwang@xidian.edu.cn; angel.rubio@mpsd.mpg.de; yugeliang@nju.edu.cn; leiwang@nju.edu.cn}

\maketitle
\section*{Abstract}
\textbf{
Electrons can form an ordered solid crystal phase ascribed to the interplay between Coulomb repulsion and kinetic energy. Tuning these energy scales can drive a phase transition from electron solid to liquid, i.e. melting of Wigner crystal. Generalized Wigner crystals (GWCs) pinned to moir\'e superlattices have been reported by optical and scanning-probe-based methods. Using transport measurements to investigate GWCs is vital to a complete characterization, however, still poses a significant challenge due to difficulties in making reliable electrical contacts. Here, we report the electrical transport detection of GWCs at fractional fillings $\nu$ = 2/5, 1/2, 3/5, 2/3, 8/9, 10/9, and 4/3 in twisted bilayer MoSe$_2$. We further observe that these GWCs undergo continuous quantum melting transitions to liquid phases by tuning doping density, magnetic and displacement fields, manifested by quantum critical scaling behaviors. Our findings establish twisted bilayer MoSe$_2$ as a novel system to study strongly correlated states of matter and their quantum phase transitions.}

\section*{Introduction}
The competition and transition between ordered and disordered phases are central to condensed matter physics and beyond. Different states of matter are often characterized by the presence or absence of localized particles, where in the latter case a liquid phase typically composed of itinerant particles is found.
This characterization famously includes metal–insulator transitions (MITs)~\cite{Tokura_MIT-review_RMP_1998} and quantum Hall states~\cite{Das_Sarma_QHE_JWS_Book_2008}. Wigner crystals and Mott insulators respectively driven by long-range Coulomb interaction~\cite{Eugene_Wigner_1934_PR, GaoXPA_Kivelson_RMP_2010} and on-site repulsion~\cite{Mott_Original+thoery_PPSA-1949} are two important examples for interaction-induced ordered phases. In principle, when quantum fluctuations are strong enough to delocalize the confined charges, the Wigner crystal will melt into a metallic phase, referred to as a quantum melting transition~\cite{Spivak_Kivelson_PRB_2004}. Similarly, varying the ratio of interaction energy to kinetic energy can drive a Mott transition~\cite{WenXP_PatrickLee_Doping-a-Mott_RMP_2006}.
Evidences of Wigner crystals have been seen in two-dimensional electron systems (2DES) at very low charge densities, including liquid helium surfaces~\cite{liquid_helium_PRL_1979}, semiconductor heterostructures~\cite{Mansour_Shayegan_WC-review_NRP_2022,Falson_and_Smet_WC-in-ZnO_NM_2022} and 2D materials~\cite{Imamoglu_Optics-in-MoSe2_Nature_2021,You_Zhou_MoSe2-BN-MoSe2_Nature_2021,Yazdani_STM-in-LL-Nature_2024}. 
However, the tunability of Wigner crystals and Mott transitions in these systems remains limited. This necessitates the development of experimental platforms that allow for systematic examinations of the phase transitions between these and other exotic states over a broader range of physical parameters.

Sparked by recent advances in graphene moir\'e systems~\cite{Dante_Review_NP_2021}, moir\'e structures utilizing  semiconducting transition-metal dichalcogenides (TMDs) have been established as highly tunable systems~\cite{Fengcheng_Wu_and_MacDonald_PRL_2018,Fengcheng_Wu_and_Das_Sarma_PRB_2020,Mak_Review_NN_2022} that are capable of capturing the physics of both Wigner crystals~\cite{Feng_Wang_GWC-and-Mott_Nature_2020,Yang_Xu_and_Mak_Optical-sensing_Nature_2020,Chenhao_Jin_and_Mak-Stipe_phase_NM_2021,Feng_Wang_STM-GWC_Nature_2021,Yongtao_Cui_MIM_GWC_NP_2021,Feldman_SET-bibiWSe2_NM_2023, Feng_Wang_STM-GWC_NN_2024,Feng_Wang_STM-Wigner-molecular-crystals_Science_2024} and Mott insulators~\cite{Lei_Wang_NM-2020,Mak_Optics-in-WSe2/WS2-Nature_2020,Imamoglu_Optics-in-MoSe2-BN-MoSe2_Nature_2020}.
The moir\'e-version of a quantum melting transition from  generalized Wigner crystals (GWCs) to liquid phases has been theoretically predicted when the bandwidth surpasses the Coulomb repulsion~\cite{KimEA_n-driven-GWC-Melting_NC_2022,MacDonald_Magnetism-and-quantum-melting-of-GWC_PRB_2023,KimEA_quantum-melting-GWC_PRL_2024}. Recently, GWCs have been observed in moir\'e TMDs using optical detections~\cite{Feng_Wang_GWC-and-Mott_Nature_2020,Yang_Xu_and_Mak_Optical-sensing_Nature_2020,Wigner-Mott-transition_NC_2022_Yanghao_Tan}, capacitance measurements~\cite{Mak_Capacitance-test-of-GWC_NN_2021}, or scanning-probe-based approaches~\cite{Feng_Wang_STM-GWC_Nature_2021,Yongtao_Cui_MIM_GWC_NP_2021}. These previous efforts focus on the rigidly crystalline regime to understand the formation of GWCs~\cite{Fengcheng_Wu_and_Das_Sarma_PRB_2020,Phillips_GWC_theory_PRB_2021,Liang_Fu_Charge-transfer_PRB_2021}, yet their quantum transitions to metallic phases have not been fully investigated. For instance, whether the quantum melting of GWCs is of first order or continuous is still under debate~\cite{MacDonald_Magnetism-and-quantum-melting-of-GWC_PRB_2023, KimEA_quantum-melting-GWC_PRL_2024, Chowdhury_Bandwidth-tuned-Wigner-transition_PRB_2022}. 
Electrical transport measurements can provide important signatures for Wigner crystals identified by the nonlinear current-voltage characteristics~\cite{Falson_and_Smet_WC-in-ZnO_NM_2022,Shayegan_Zero-B-WC-in-AlAs_PNAS_2020} and furnish us to examine the resistance scaling behaviors in the phase transition process~\cite{Dobrosavljevic-Wigner-Mott-scaling_PRB_2012}.  
Although transport experiments have shown continuous Mott transitions in twisted p-type 2D semiconductor WSe$_2$~\cite{Lei_Wang_Quantum-criticality_Nature_2021} and aligned heterobilayer WSe$_2$/MoTe$_2$~\cite{Mak_Scaling_Nature_2021},the observation and quantum melting of GWCs have not yet been observed. 

In this work, we fabricate AA-stacked twisted bilayer MoSe$_2$ (tMoSe$_2$), an n-type TMD which allows us to access the moir\'e physics on the rarely-studied electron-doped side. While the splitting of the conduction band states due to spin-orbit coupling (SOC) in n-type TMDs is reduced as compared to the valence band states, the effect of SOC on the moir\'e physics is expected to be different~\cite{Guangyu_Zhang_tMoS2_PRL_2023,Shengjun_Yuan-SOC-in-tMOS2_PRB_2020}.}
Using transport measurements, we observe generalized Wigner crystal states at multiple fractional electron fillings, together with a magnetic field stabilized Mott state at $\nu$ = 1. The quantum melting of these GWCs states can be finely controlled by tuning electric field, magnetic field or electron density, whereas the Mott state behaves markedly differently when changing these parameters.

\begin{figure*}[t]
\begin{center}
\includegraphics[width=1\linewidth]{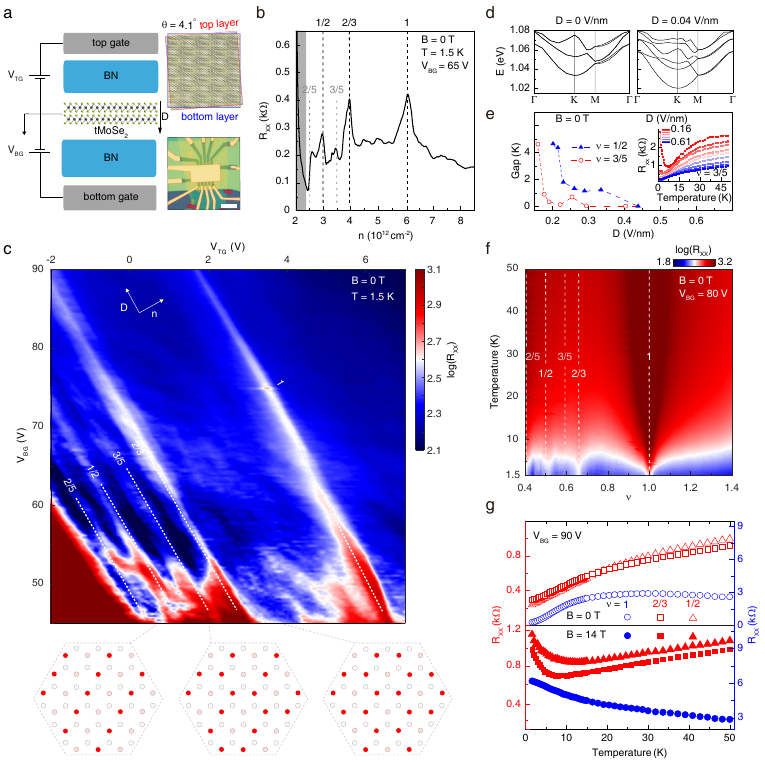}
\caption{   
\textbf{Emergence of correlated states at integer and fractional fillings in tMoSe$_2$.} 
        \textbf{a,} Schematic side view (left) and optical image of the device (lower right, the scale bar is 10 $\mu$m) and depiction of the twisted bilayer MoSe$_2$ moir\'e superlattice (upper right). The sign of the displacement field, $D$, is defined to be positive when the field points from the back gate to the twisted sample. 
        \textbf{b,} Longitudinal resistance $R_{\textrm{xx}}$ plotted against carrier density, $n$, at $V_{\textrm{BG}}$ = 65 V and $T$ = 1.5 K. The dashed lines mark the moir\'e filling factor ($\nu$) of each resistive peak. 
        \textbf{c,} Color map of $R_{\textrm{xx}}$ as a function of $V_{\textrm{TG}}$ and $V_{\textrm{BG}}$ under zero perpendicular magnetic field ($B$ = 0 T) at 1.5 K. Proposed charge orders for $\nu$ = 1/2 (left), 3/5 (middle) and 2/3 (right) states are listed below.
        \textbf{d,} DFT calculation of hybridized band structures of 4$^\circ$ tMoSe$_2$ with and without $D$. 
        \textbf{e,} Evolution of charge gaps with $D$ for $\nu$ = 1/2 (blue solid triangle) and 3/5 (red open circle) states. The inset shows typical temperature-dependent $R_{\textrm{xx}}$ of the $\nu$ = 3/5 state with varying $D$.
        \textbf{f,} 2D map of $R_{\textrm{xx}}$ as a function of filling factor $v$ and temperature $T$ at $V_{\textrm{BG}}$ = 80 V and $B$ = 0 T.
        \textbf{g,} Temperature-dependent $R_{\textrm{xx}}$ measured for $B$ = 0 T (hollow symbols, upper) and 14 T (solid symbols, lower) at $V_{\textrm{BG}}$ = 90 V, respectively. $\nu$ = 1/2, red triangle (left axis); $\nu$ = 2/3, red square (left axis); $\nu$ = 1, blue circle (right axis). 
}

\label{fig:fig1}
\end{center}
\end{figure*}

\section*{Results and Discussion}
\noindent\textbf{Correlated states at integer and fractional fillings}

Fig.~\ref{fig:fig1}a shows our dual-gated device structure. Importantly, bismuth is used as the metallization layer to achieve good ohmic contacts~\cite{Jing_Kong_Bi-contact_Nature_2021} (see Methods). We use polarization-dependent second-harmonic generation (SHG) to determine a twist angle of 4.1$^\circ$ (see Extended Data Fig. 1). 
Accordingly, $n_\textrm{s}$, the electron density of half-band filling, is calculated to be $\sim6.0\times 10^{12}$ cm$^{-2}$. The moir\'e filling factor is defined as $\nu$ = $n/n_\textrm{s}$, where $n$ represents the gate-induced electron density. The dual-gate device geometry allows independent control over the displacement field, $D$, and electron density, $n$, via top and back gate voltages ($V_{\textrm{TG}}$ and $V_{\textrm{BG}}$, respectively). Fig. \ref{fig:fig1}b shows the longitudinal resistance $R_{\textrm{xx}}$ as a function of $n$ at $V_{\textrm{TG}}$ = 65 V and $T$ = 1.5 K, in which pronounced resistance peaks can be observed at $\nu$ = 1, 2/3 and 1/2, corresponding to the commensurate fillings of one electron per moir\'e unit cell, two electrons per three moir\'e cells, and one electron per two moir\'e unit cells, respectively (See Extended Data Fig. 2 for similar data observed in another device). In addition, relatively weaker resistance peaks at $\nu$ = 2/5 and 3/5 are also found. Fig. \ref{fig:fig1}c shows the color map of $R_{\textrm{xx}}$ versus $V_{\textrm{TG}}$ and $V_{\textrm{BG}}$, exhibiting features of high resistance traces along constant multiple fractional or integer filling lines. Analogous fractional filling states have not been reported so far in hole-doped moir\'e TMDs using transport measurement in spite of the observations through  optical sensing technique~\cite{Yang_Xu_and_Mak_Optical-sensing_Nature_2020} or microwave impedance microscopy~\cite{Yongtao_Cui_MIM_GWC_NP_2021}.

To connect to these experimental findings, we perform density functional theory (DFT) calculations. The electronic band structure in Fig.~\ref{fig:fig1}d confirms the presence of a miniband on the electron side with a bandwidth of $\sim$70 meV for a twist angle close to the experimental one under zero (left panel) and large (right panel) $D$ (see Supplementary Materials for details). We find that the lowest-lying conduction bands are faithfully represented by two localized Wannier orbitals centered at the Se/Mo (Mo/Se) stacking position of the moir\'e unit cell, which we use to address the impact of long-ranged Coulomb interactions on the electronic ordering tendencies of the system on the Hartree-Fock level (see Supplementary Materials for details). It is found that the longer ranged interactions are particularly important at fractional fillings and can stabilize insulating phases with modulated charge patterns on the moir\'e honeycomb lattice as proposed in the lower cartoons of Fig.~\ref{fig:fig1}c. For $\nu$ = 1/2, the electrons form a stripe order, and for $\nu$ = 2/3, the electrons occupy a honeycomb lattice, while for $\nu$ = 3/5, the electrons occupy an elongated honeycomb lattice. Recently, such correlated states occurring at fractional fillings of moir\'e superlattices are referred to as generalized Wigner crystals~\cite{Feng_Wang_GWC-and-Mott_Nature_2020,Yang_Xu_and_Mak_Optical-sensing_Nature_2020,Feng_Wang_STM-GWC_Nature_2021,Yongtao_Cui_MIM_GWC_NP_2021}. 

Experimentally, we extract the charge gaps of $\nu$ = 3/5 and 1/2 states from thermal activations in Fig.~\ref{fig:fig1}e (see Extended Data Fig.~3). In the absence of magnetic fields, the maximal charge gap reaches values of 2.2 K (at $D$ = 0.11 V/nm), 4.6 K (at $D$ = 0.16 V/nm) and 5.2 K (at $D$ = 0.20 V/nm) for $\nu$ = 2/3, 3/5 and 1/2 states, respectively. We further identify these fractional filling states as GWCs by measuring the differential resistance, $dV/dI$, as a function of the applied DC current, $I_{\textrm{DC}}$ (see Extended Data Fig.~4). The differential resistances display strong nonlinear features with pronounced peaks centered at zero $I_{\textrm{DC}}$. At a low bias, electrons are localized by the moir\'e superlattice. A bias above a sufficiently high threshold can lead to a sharp decrease in differential resistance, which is often referred to as the depinning~\cite{Shayegan_Zero-B-WC-in-AlAs_PNAS_2020} or onset of sliding of Wigner crystals~\cite{Shayegan2023sliding}.\\

\begin{figure*}[t]
\begin{center}
\includegraphics[width=1\linewidth]{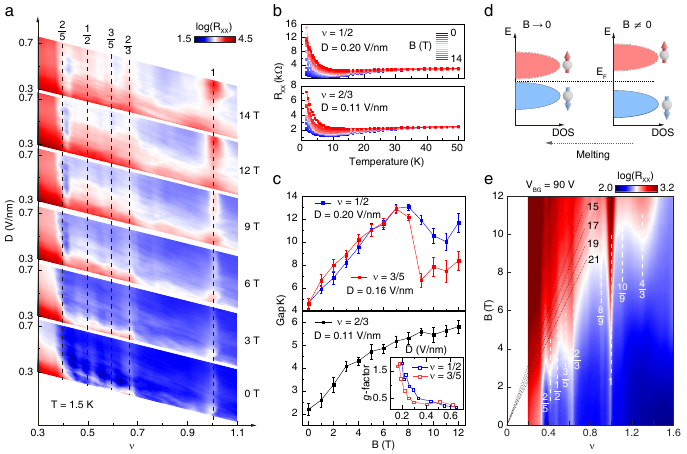}
\caption{
    \textbf{Magnetic and displacement field tuning of generalized Wigner crystals.} 
    \textbf{a,} $R_{\textrm{xx}}$ as a function of $D$ and $\nu$ under $B$ = 0, 3, 6, 9, 12 and 14 T at 1.5 K (from down to up).
    \textbf{b,} Temperature dependence of $R_{\textrm{xx}}$ under different $B$ for $\nu$ = 1/2 (upper penal) and 2/3 (lower penal) states.
    \textbf{c,} Charge gap evolution with $B$ for $\nu$ = 2/3 (black), 3/5 (red) and 1/2 (blue) states. The inset shows the variation of $g$-factor with $D$. Error bars reflect the uncertainty in determining the charge gaps.
    \textbf{d,} The schematic of the transition for generalized Wigner crystal states in a magnetic field.
    \textbf{e,} $\nu-B$ resistance map of $R_{\textrm{xx}}$ at 1.5 K and $V_{\textrm{BG}}$ = 90 V. The black dotted lines label the dominant sequence in the Landau fan projecting to the band edge of tMoSe$_2$. The white dashed lines indicate the emergent correlated states.
    }

\label{fig:fig2}
\end{center}
\end{figure*}

\noindent\textbf{Magnetic and displacement field effects} 

Next, we investigate the response of the correlated states under the influence of external magnetic and displacement fields. By increasing $D$, the correlated states at fractional fillings turn metallic as the bands become more dispersive, see Fig. \ref{fig:fig1}d. The inset in Fig.~\ref{fig:fig1}e shows a representative $D$-induced MIT at $\nu$ = 3/5. Similar data for the $\nu$ = 2/3 and 1/2 states are provided in Extended Data Fig.~3. Next, we map out $R_{\textrm{xx}}$ versus $(\nu, T)$ under a large $D$ ($V_{\textrm{BG}}$ = 80 V) as shown in Fig.~\ref{fig:fig1}f. The resistance of all correlated states at fractional and integer fillings increases as a function of temperature as explicitly shown for $\nu=1$, 2/3, and 1/2 in the upper penal of Fig.~\ref{fig:fig1}g. However, the insulating behavior is recovered in the presence of a strong magnetic field ($B = 14$ T) as demonstrated in the lower panel of Fig.~\ref{fig:fig1}g. No response to in-plane magnetic field can be observed due to Ising-type SOC~\cite{Lei_Wang_NM-2020,Mak_Scaling_Nature_2021} (see Extended Data Fig.~5).

Fig. \ref{fig:fig2}a shows a series of color maps of $R_{\textrm{xx}}$ for different magnetic fields in dependence of $D$ and $\nu$ at $T=$ 1.5 K. Both the fractional filling and $\nu=1$ states become more resistive when increasing $B$.
The temperature dependence of $R_{\textrm{xx}}$ depicted in Fig. \ref{fig:fig2}b for the states at $\nu$ = 1/2 and 2/3 unambiguously shows that the application of a magnetic field strengthens the insulating behavior. To quantify this, we extract the magnetic field-dependent charge gaps in Fig. \ref{fig:fig2}c, from which we can also extract the effective $g$-factor of the generalized Wigner crystal states. Interestingly, the $g$-factor decreases for increasing displacement field (bottom panel inset), taking values of $g=$ 1.6 (0.3) at displacement fields of $D$ = 0.16 (0.65) V/nm. 

Multiple possible effects can strengthen the insulating behavior with applying a magnetic field~\cite{pippard1989magnetoresistance,goldman1990evidence,chen2023magnetic,cao2020tunable}. In our system, the behavior of GWCs is qualitatively akin to those of the correlated states with the spin-polarized nature in other moir\'e materials~\cite{cao2020tunable,liu2020tunable,shen2020correlated,Feldman_SET-bibiWSe2_NM_2023}. To explain such effect, we consider the effect of Zeeman splitting as illustrated in Fig.~\ref{fig:fig2}d. Zeeman coupling causes an energetic splitting of the spin species by an energy $E_{\mathrm Z} = g \mu_B B$. 
Therefore, the spin-polarized correlated insulators will experience an increase of their gap when applying a magnetic field. When decreasing $D$, correlation effects are stronger due to the reduced bandwidth, as shown in the DFT calculations (Fig.~\ref{fig:fig1}d). Therefore, the fractional filling states have a stronger tendency of spontaneous spin-polarization induced by correlations. Fig.~\ref{fig:fig2}e shows the Landau fan diagram at $V_{\textrm{BG}}$ = 90 V,which reveals a two-fold degenerated Landau levels stemming from the conduction band minimum at K-point~\cite{Tutuc_MoSe2_PRB_2018}. We observe that the $B$-field induced enhancement is found for all GWCs as well as $\nu$ = 1 state, which is unable to be predicted by other enhancement mechanisms ~\cite{pippard1989magnetoresistance,goldman1990evidence,chen2023magnetic}. Strikingly, additional correlated states emerge at larger magnetic fields for fractional fillings of $\nu=$ 4/3, and more faintly even 8/9 and 10/9.
To fully confirm the nature of their magnetic orders, other optical techniques need to be used and will be reported in our future work.\\

\begin{figure*}[t]
\begin{center}
\includegraphics[width=1\linewidth]{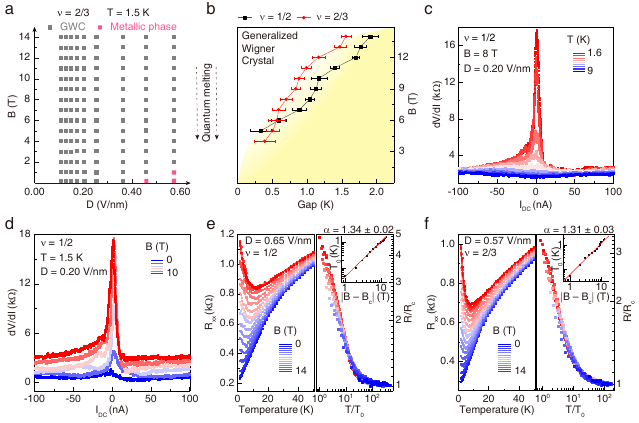}
\caption{
    \textbf{Melting transitions of generalized Wigner crystal states.} 
    \textbf{a,} Phase diagram in $B-D$ space for $\nu$ = 2/3 state at $T$ = 1.5 K. GWC stands for generalized Wigner crystal.
    \textbf{b,} Evolution of charge gap with $B$ for $\nu$ = 1/2 ($D$ = 0.65 V/nm) and $\nu$ = 2/3 ($D$ = 0.57 V/nm) states.
    The yellow shaded region represents the metallic phase. Error bars reflect the uncertainty in determining the charge gaps.
    \textbf{c,} $dV/dI$ as a function of $I_{\textrm{DC}}$ for $\nu$ = 1/2 state with different temperatures under $B$ = 8 T and $D$ = 0.20 V/nm.
    \textbf{d,} $dV/dI$ as a function of $I_{\textrm{DC}}$ for $\nu$ = 1/2 state with varied $B$ under $T$ = 1.5 K and $D$ = 0.20 V/nm.
    \textbf{e,} Temperature dependence of $R_{\textrm{xx}}(T)$ under varied $B$ for $\nu$ = 1/2 state (left) and the corresponding scaling plot of the normalized resistivity $R_{\textrm{xx}}(T)/R_\textrm{c}$ versus $T/T_0$ (right). The inset shows the corresponding $T_0$ vs $|B - B_\textrm{c}|$.
    \textbf{f,} Scaling analysis for $\nu$ = 2/3 state. The determination of scaling parameter $T_0$, critical magnetic field $B_\textrm{c}$ and critical resistance $R_\textrm{c}$ are discussed in the main text. The $D$ value in c-f is specially selected at each fractional filling to ensure the observation of melting transitions.   
}

\label{fig:fig3}
\end{center}
\end{figure*}

\noindent\textbf{Quantum melting transitions}

We examine the temperature dependence of $R_{\textrm{xx}}$ to determine the metallic or insulating phases in the $B-D$ space (see Extended Data Fig.~6 and 7). As shown in Fig.~\ref{fig:fig3}a, the insulating phase occupies most of the region in the accessible $B-D$ range for the $\nu$ = 2/3 state. The $\nu$ = 1/2 state behaves similarly. We show in the Supplemental Material that the behavior of the two phases is consistent with our mean-field analysis (see Extended Data Fig. 8). Within the insulating regime, Fig. \ref{fig:fig3}c displays $dV/dI$ as a function of $I_{\textrm{DC}}$ for the $\nu$ = 1/2 state at various temperatures. The sharp peak in $dV/dI$ indicates a nonlinear transport at low temperatures, from which the thermal melting can be seen through the collapse of the peak. The  insulating states at $\nu$ = 1/2, 2/3 and 3/5 can be melted thermally at $T\approx$ 5 K as indicated by the temperature-dependent $dV/dI$ curves (see Extended Data Fig. 9). Besides thermal melting, through tuning other parameters, the GWCs can also melt to a metallic phase at base temperature in the large $D$ and small $B$ region. 

In the following, we focus on the quantum melting transition triggered by decreasing the magnetic field as indicated by the dashed arrow between Fig.~\ref{fig:fig3}a and b. We show the extracted charge gaps at $\nu=$ 1/2 and 2/3 for different applied magnetic fields in Fig.~\ref{fig:fig3}b. As $B$ is decreased, the charge gap closes and a metallic state emerges. Accordingly, the $dV/dI$ peak vanishes as shown in Fig.~\ref{fig:fig3}d.
When $D$ is decreased or the electron density deviates from $\nu$ = 1/2, the nonlinear transport behavior also nearly disappears (see Extended Data Fig. 10). The correlated states at $\nu$ = 2/3 and $\nu$ = 3/5 show similar characteristics (see Extended Data Fig.~11). Therefore, our data indicates that the GWCs transit into a liquid phase due to quantum melting via decreasing $B$, increasing $D$ or moving away from densities at fractional fillings.

We further analyze the temperature dependent behavior of $R_{\textrm{xx}}$ in the magnetic field-driven quantum phase transitions shown in Fig.~\ref{fig:fig3}e and f. 
We collapse the $R_{\textrm{xx}}$ curves by scaling with the critical resistance $R_\textrm{c}(T)$ of the critical magnetic field ($B_\textrm{c}$ = 0 T), by the form $R_{\textrm{xx}}(T) = R_\textrm{c}(T)f(T/T_0(\delta B))$~\cite{Dobrosavljevic_Mott-Scaling_PRB_2013}, with $T_0(\delta B) \sim |\delta B|^{z\nu}$ and $\delta B = B - B_\textrm{c}$. The fitting parameter $T_0$ turns out to be comparable to the magnitude of the charge gap. All the data points for $R_{\textrm{xx}}(T)/R_\textrm{c}(T)$ against $T/T_0$ collapse well onto one curve, with a critical exponent of $z\nu$ = 1.34 $\pm$ 0.02 and 1.31 $\pm$ 0.03 for $\nu$ = 1/2 and 2/3 states, respectively. The scaling behavior at $\nu$ = 3/5 state is comparable (see Extended Data Fig. 12). It is noted that a similar power law with an exponent $\sim$ 1.6 was reported for the scaling of a carrier-density-driven MIT in silicon-based 2DES~\cite{kravchenko1995scaling}.\\

\begin{figure*}[t]
\begin{center}
\includegraphics[width=1\linewidth]{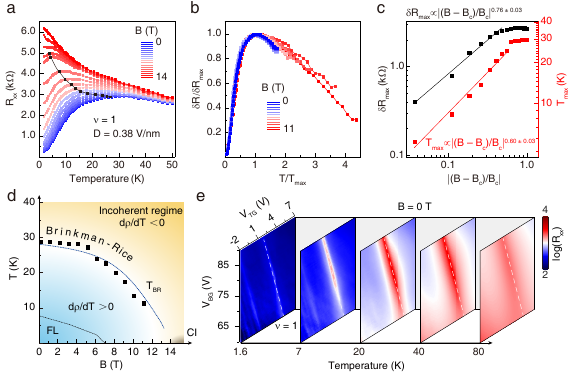}
\caption{
    \textbf{Phase transitions of $\nu$ = 1 state.}
    \textbf{a,} Temperature dependence of $R_{\textrm{xx}}$ at $D$ = 0.38 V/nm for various magnetic field $B$. 
    \textbf{b,} The scaling plot of the ratio $\delta R(T)$/$\delta R_{\textrm{max}}$ as a function of $T/T_{\textrm{max}}$.
    \textbf{c,} Plots of $T_{\textrm{max}}$ and $\delta R_{\textrm{max}}$ as a function of reduced magnetic field [$(B - B_\textrm{c})/B_\textrm{c}$].
    \textbf{d,} Electronic transport regimes around the MIT of $\nu$ = 1 state. CI stands for correlated insulator and FL stands for Fermi liquid.
    \textbf{e,} $R_{\textrm{xx}}$ versus $V_{\textrm{BG}}$ and $V_{\textrm{TG}}$ at multiple temperatures under zero magnetic field. The white dashed lines indicate the $\nu$ = 1 state.
}

\label{fig:fig4}
\end{center}
\end{figure*}

\noindent\textbf{Phase transition of $\nu$ = 1 state}

We next investigate the phase transition of $\nu$ = 1 state. Fig.~\ref{fig:fig4}a shows the temperature dependent $R_{\textrm{xx}}(T)$ of $\nu$ = 1 state at different $B$. When $B \langle 13$ T, the curves indicate metallic behavior ($dR_{\textrm{xx}}$/$dT \rangle 0$) below a temperature $T_{\textrm{max}}$, while above $T_{\textrm{max}}$ the behavior turns insulating-like ($dR_{\textrm{xx}}$/$dT \langle 0$), resulting in a resistivity maximum $R_{\textrm{max}}$ at $T_{\textrm{max}}$. Such a pronounced $R_{\textrm{max}}$ has been observed in 2DES~\cite{Dobrosavljevic-Wigner-Mott-scaling_PRB_2012}, heavy fermion systems~\cite{Heavy-fermion-CeCu6_PRB_1985} and charge-transfer organic salts~\cite{Organic-Mott_PRL_2015}, in which the essential mechanism of transport relies on thermal destruction of coherent quasi-particles due to strong inelastic electron-electron scattering. At $B$ = 0 T, the Fermi-liquid behavior is restricted to a very limited range, $T \ll T_{\textrm{max}} \ll T_{\textrm{F}}$, where $T_{\textrm{max}}$ is the coherence temperature and $T_{\textrm{F}}$ is the Fermi temperature. As the temperature increases, the electron mean free path becomes comparable to or smaller than the moir\'e wavelength, leading to incoherent transport. When the electron spin is taken into account, the coherent quasiparticles can also be destroyed by adding a large enough Zeeman energy $E_\textrm{Z}$ to open a small band gap in the miniband with a bandwidth of $W_0$ ($g\mu_\textrm{B}B + W_0 \geq W^*$, $W^*$ represents the bandwidth of metallic phase at the critical point of MITs). This is in line with our experimental observations that $T_{\textrm{max}}$ decreases but $R_{\textrm{max}}$ increases with magnetic field. Thus, we argue that $T_{\textrm{max}}$ is the coherence temperature beyond which incoherent transport sets in. Fig. \ref{fig:fig4}d displays a proposed phase diagram for $\nu$ = 1 state. On the metallic side of the diagram, resistive maxima at the Brinkman-Rice temperature $T_{\textrm{BR}}=T_{\textrm{max}}$ signal the destruction of resilient quasiparticles and the crossover to an incoherent transport regime~\cite{Dobrosavljevic_Phase-diagram_NC_2021}.

Next, we perform a scaling analysis using $\delta R(T)$ = $\delta R_{\textrm{max}}f(T/T_{\textrm{max}})$, where $\delta R(T)$ = $R(T) - R_0$ and $\delta R_{\textrm{max}} = R_{\textrm{max}} - R_0$. We assume that the resistance follows the formula $R(T)$ = $R_0$ + $\delta R(T)$. Here, $R_0$ is the residual resistivity, and $\delta R(T)$ is the temperature dependent resistance dominated by inelastic electron-electron scattering. As shown in Fig. \ref{fig:fig4}b, the data points over a wide $B$ range collapse onto a single curve. Both $T_{\textrm{max}}$ and $\delta R_{\textrm{max}}$ show a power-law dependence on the reduced magnetic field [$(B - B_\textrm{c})/B_\textrm{c}$] with an exponent of 0.60 and 0.76 ($B_\textrm{c}$ = 13.5 T) as shown in Fig. \ref{fig:fig4}c. Similar values are found for 2DES when scaling with charge density~\cite{Dobrosavljevic-Wigner-Mott-scaling_PRB_2012}. Besides tuning the magnetic field, a MIT with perfect scaling behavior is also found when altering the density near $\nu=1$ (see Extended Data Fig. 13).
The $R_{\textrm{xx}}(T)$ curves for $\nu$ = 1 state change from quadratic to linear behavior as $D$ decreases despite the absence of a MIT (see Extended Data Fig. 14), implying that the system is approaching the critical point of a quantum phase transition.

The physics of $\nu$ = 1 state can also be interpreted in terms of a Pomeranchuck effect describing a transition from a low-entropy electronic liquid to a high-entropy correlated state~\cite{Pomeranchuk-effect-original-paper_1950,Yang_Xu_3L-MoTe_2/WSe2_PRX_2022}. Fig.~\ref{fig:fig4}e shows the dual-gate maps of $R_{\textrm{xx}}$ at different temperatures. In marked contrast to the gradual disappearance of fractional filling insulators, the resistance of the $\nu$ = 1 state initially increases up to $\sim$30 K and subsequently attenuates (see Extended Data Fig. 15). 

In conclusion, our observations suggest that a continuous quantum phase transition can be achieved in generalized Wigner crystal states as well as Mott state in tMoSe$_2$ by carefully tuning a multi-parameter space consisting of magnetic field, carrier density, and displacement field. Our system, which hosts both Wigner and Mott states in one single material, can serve as an ideal platform to continuously tune local and long-range Coulomb interactions. The rich phenomenology combined with multiple degrees of experimental control in tMoSe$_2$ would contribute to unraveling novel states of matter and provide further insights into the quantum phase transitions between ordered and disordered correlated states.

\bigskip

\section*{Methods}
\noindent\textbf{Sample fabrication}

The tMoSe$_2$ devices were fabricated using the pick-up transfer~\cite{Lei_Wang_Science_2013}. Both monolayer MoSe$_2$ and h-BN flakes were mechanically exfoliated onto SiO$_2$/Si substrates. Then the monolayer MoSe$_2$ flake was cut into two pieces with an AFM tip. All stacks were assembled utilizing a polypropylene carbonate (PPC)/polydimethylsiloxane (PDMS) stamp on a glass slide~\cite{Lei_Wang_NM-2020,Lei_Wang_Quantum-criticality_Nature_2021}. For the subsequent metallization process, the PPC/hBN/tMoSe$_2$ stack was flipped and placed on another SiO$_2$/Si substrate with tMoSe$_2$ as the top layer. A thin layer of h-BN flake (5-8 nm) was patterned into a Hall bar geometry by CHF$_3$/O$_2$ etching, picked up with a new PPC/PDMS holder, dropped onto the surface of tMoSe$_2$ and annealed in ultrahigh vacuum. Bi/Au (6/10 nm) metal layer was deposited on tMoSe$_2$ across the prepatterned h-BN using electron-beam lithography and electron-beam evaporation to form the metal/semiconductor contacts~\cite{Jing_Kong_Bi-contact_Nature_2021}. Another h-BN flake (20-30 nm) was picked up and released on the as-deposited Bi/Au electrode. Finally, electron-beam lithography and evaporation were performed again to deposit a Cr/Pd/Au (5/15/100 nm) metal layer to connect the Bi/Au electrodes and simultaneously define the top gate of the Hall bar channel. The technical details of device fabrication herein will be reported elsewhere.\\ 

\noindent\textbf{Twist angle determination}

Optical SHG and atomic force microscopy were used to roughly measure the twist angle $\theta$ between the top and bottom MoSe$_2$. The value of $\theta$ is further confirmed with the half filling density $n_\textrm{s}$ = 8(1$- \textrm{cos}\theta$)/$\sqrt{3}a^2$ of the moir\'e superlattices~\cite{Lei_Wang_NM-2020}, where $a$ = 0.3288 nm is the in-plane lattice constant of H phase MoSe$_2$. In detail, $n_\textrm{s}\sim6.0\times 10^{12}$ cm$^{-2}$ can be extracted from the fan diagram, matching well with the twist angle extracted from SHG.\\ 

\noindent\textbf{Transport measurements}

Electrical transport measurements were carried out in a dilution refrigerator with a base temperature of 1.5 K and a maximum magnetic field of 14 T. All the data in this work were obtained using the standard low-frequency lock-in technique at an excitation frequency of 17.777 Hz with an a. c. current of 20 nA. The four-terminal resistance was acquired by recording the source–drain current and the four-probe voltage concurrently with two lock-in amplifiers.

\bigskip

\section*{Data availability}
The data that support the findings of this study are available from the corresponding authors upon request.

\bibliography{Refs.bib}

\bigskip

\section*{Acknowledgments}
L.W. acknowledges the National Key Projects for Research and Development of China (Grant Nos. 2022YFA1204700 and 2021YFA1400400), Program for Innovative Talents and Entrepreneur in Jiangsu (Grant No.JSSCTD202101), Natural Science Foundation of Jiangsu Province (Grant Nos. BK20220066 and BK20233001) and Nanjing University International Research Seed Fund. H.W. acknowledges support from National Natural Science Foundation of China (Grant No. 61804117), Natural Science Foundation of Shaanxi Province (Grant No. 2022JM-364) and Guangdong Basic and Applied Basic Research Foundation (Grant No. 2022A1515111075). K.W. and T.T. acknowledge support from the JSPS KAKENHI (Grant Nos. 21H05233 and 23H02052) and World Premier International Research Center Initiative (WPI), MEXT, Japan. D.A.R. and Y.H. acknowledge support from the University of Wisconsin-Madison, Office of the Vice Chancellor for Research and Graduate Education with funding from the Wisconsin Alumni Research Foundation.

\bigskip

\section*{Author Contributions Statement}
L.W. conceived and designed the experiment. Q.J.Z., Q.Z., H.W. and G.Y. fabricated the samples. Q.J.Z. and H.W. performed the transport measurements. L.W., H.W. and Q.J.Z. analyzed the data. Y.H. and D.A.R. grew the MoSe$_2$ crystal. K.W. and T.T. grew the h-BN crystal. L.X. and Q.X. performed the density functional theory calculation. A.R., D.M.K., X.C. and A.F. performed the mean field theory calculation. L.W., H.W. and Q.J.Z. wrote the manuscript with input from all the authors.

\bigskip

\section*{Competing Interests Statement}
The authors declare no competing interest.

\end{document}